\documentclass[%
 amsmath,amssymb,
 aps,
 reprint,
nofootinbib
]{revtex4-1}
\usepackage[dvipsnames]{xcolor}

\usepackage{graphicx}
\usepackage{hyperref}
\hypersetup{colorlinks = true, linkcolor = [rgb]{0.19411,0.51882,0.667058}, urlcolor = [rgb]{0.125490,0.29542,0.1647058}, citecolor = [rgb]{0.75882,0.37411,0.14117}}

\usepackage{quantum}
\usepackage{bm}

\begin{document}

\title{Quantum Conference Key Agreement with Photon Loss}

\author{Phattharaporn Singkanipa}
\author{Pieter Kok}\email{p.kok@sheffield.ac.uk}
\affiliation{Department of Physics and Astronomy, The University of Sheffield, Sheffield, S3 7RH, UK}

\begin{abstract}\noindent
Conference key agreement (CKA) is an information processing task where more than two parties want to share a common secret key. Here, we present a loss-resilient protocol for CKA, based on redundant encoding and error correction. Our protocol provides a speed-up in transmission rate over the existing lossy CKA protocol. However, encoding and error correction come with extra cost. We show that, using photon sources with creation probability $p \gtrsim 0.3$, our protocol's secret key rate can overcome the existing protocol's. Hence, high probability entangled photon sources are required for realistic implementation of our loss-resilient protocol.
\end{abstract}

\maketitle


\section{Introduction} \label{sec:intro}
\noindent
Quantum communication promises to provide better security \cite{Wootters:1982zz,Bennett_2014,PhysRevLett.67.661, Zhang_2017} and better speed \cite{grover1996fast,Shor_1997} using fewer resources \cite{Harrow_2004} than classical communication. For example, it  uses quantum state collapse when a measurement is made to detect the presence of eavesdroppers on communication channels, and entanglement to increase the efficiency of information transfer \cite{Bennett93}. The security of quantum key distribution was proven to derive from the laws of quantum mechanics \cite{mayers2001unconditional,lo1999unconditional,shor2000simple}. This requires that the parties can authenticate each other, i.e., it is assumed that the eavesdropper is unable to pretend to be one of the communicating parties. There are many protocols to enhance security in quantum communication to achieve this unconditional security, e.g., BB84 \cite{Bennett_2014}, Ekert 91 \cite{PhysRevLett.67.661} and Bennett 92 \cite{PhysRevLett.68.3121}. They slightly reduce the rate of information transmission to achieve unconditional security by sacrificing a subset of the shared bit string to detect eavesdroppers.

Quantum Conference Key Agreement (CKA) is an entanglement-assisted protocol that allows $N$ parties to establish a secret key efficiently. The two most common ways to share entanglement between $N$ parties are (1) to share bipartite entanglement between all pairs among the communicating parties, and (2) to share a single $N$-partite entangled state at once. By considering the achievable channel capacities for the two methods, it was proved that the latter method is more efficient than the former \cite{murta2020quantum,Epping_2017,Grasselli_2018,Ribeiro_2018}. To ensure the security of CKA, the BB84 protocol is extended to work between more than two parties, known as $N$-BB84 \cite{Grasselli_2018}. Originally, the CKA protocol shares an $N$-partite Greenberger-Horne-Zeilinger (GHZ) state once per round. To incorporate $N$-BB84 into CKA, the GHZ state is shared for $L$ rounds. After receiving a qubit in each round, there are two types of actions for each party to perform, called type-1 and type-2. Type-1 rounds are used to construct the secret key for CKA. They require each party to perform a measurement in the Pauli $Z$-basis $\{\ket{0},\ket{1}\}$. Type-2 rounds require each party to perform measurement in X-basis $\{\ket{+},\ket{-}\}$. The measurement results are used to quantify noise. There are $m=pL$ of type-2 rounds, where $p$ is a probability chosen to optimise security and secret key rate. The measurement results of the $L-m$ type-1 rounds are used to construct the secret key. To determine when the parties perform a type-2 round, another secret bit string is required. This way, even if an eavesdropper intercepts a different party in each round, she has no knowledge whether the round is type-1 or type-2. This ensures the security of the protocol \cite{Grasselli_2018}.

\begin{figure}[b!]
\centering
\includegraphics[width=0.65\columnwidth]{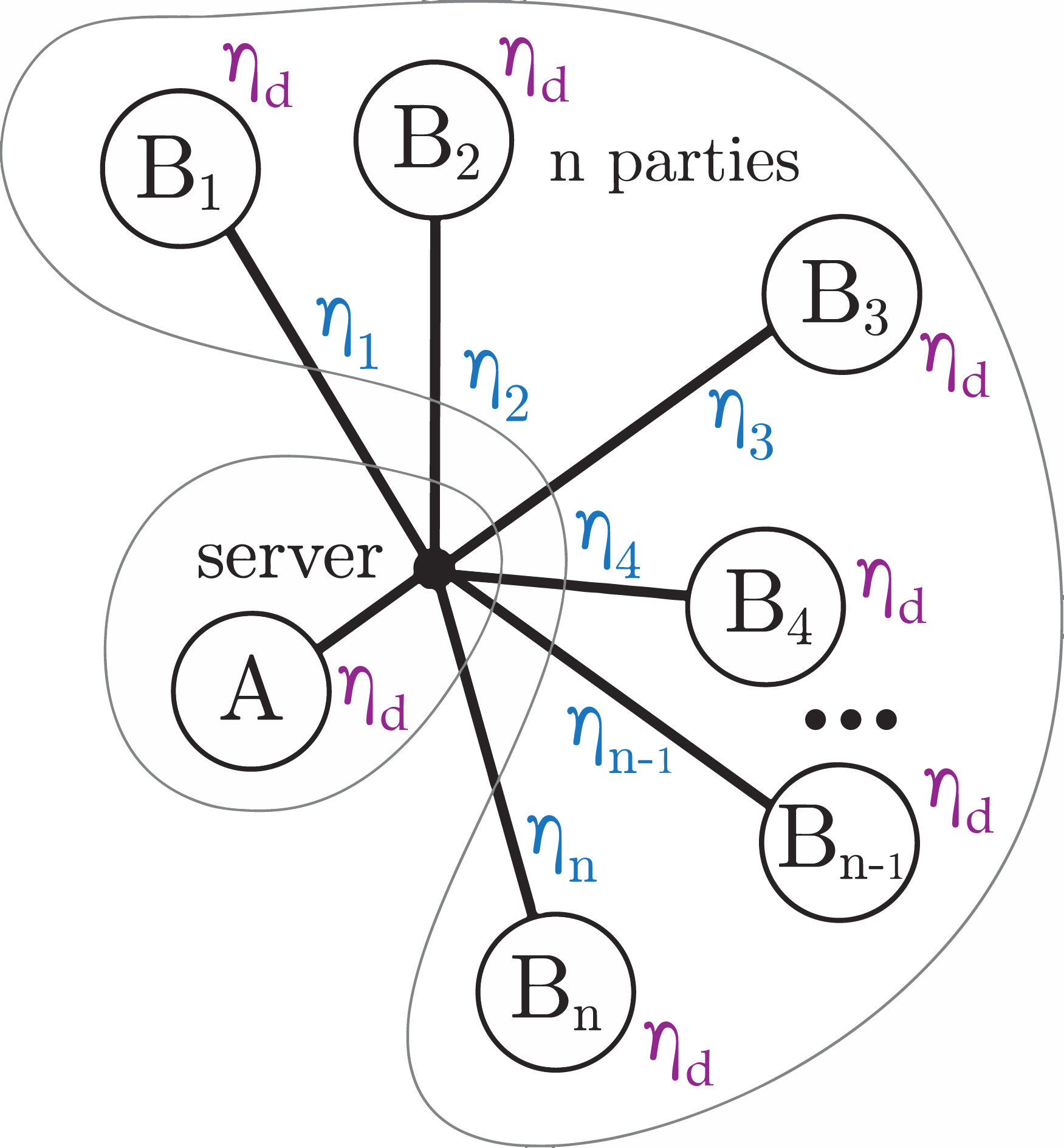}
\caption{Schematic diagram of quantum conference key agreement between $N=n+1$ parties, $A, B_1,\dots, B_n$. The server produces and transmits entangled photons to each party. Each party $B_i$ has probability $\eta_i$ of losing a photon along the transmission line. Party $A$ is very close to the server and is assumed to have no transmission loss. Each party is assumed to have the same loss per photon $\eta_d$ in the photodetectors.}
\label{setupeta}
\end{figure}

In this paper, we consider the practical implementation of CKA using distributed photons, and study the effect of lossy transmission channels. A diagram for $N$-partite entanglement-assisted CKA is shown in Fig.~\ref{setupeta}. A central server,  co-located with the communicating party $A$, produces and transmits entangled photons to the parties $A$, $B_1,B_2,\dots,B_n$. The parties may be at varying distances from the server, and assuming fibre-optical transmission cables, the resulting photon losses $\eta_j$ will generally be different for different parties. In our analysis we assume that party $A$ is so close to the server that its fibre losses are negligible. We consider $N$-partite photonic GHZ states
\begin{align}\label{eq:mngero}
 \ket{\text{GHZ}_N} = \frac{ \ket[A]{H} \ket[B_1]{H}..\ket[B_n]{H} +   \ket[A]{V}\ket[B_1]{V}..\ket[B_n]{V}}{\sqrt{2}}\, ,
\end{align}
where $\ket[j]{H}$ and $\ket[j]{V}$ denote a horizontally and vertically polarised photons, respectively, received by party $j$. It is well-known that the GHZ state is very sensitive to photon loss. If even one photon is lost anywhere in the protocol, the protocol fails and has to be attempted again. In this paper, we consider the photon loss from transmission lines and photodetectors in detail, and explore how error correction protocols such as parity encoding and redundant encoding can be used to protect the fragile GHZ states. This will place strong requirements on the entanglement generation sources. 

This paper is organised as follows: in Sec.~II we review the preparation of photonic GHZ states. In Sec.~III we consider photon loss and error correction. In Sec.~IV we present a loss tolerant protocol for conference key agreement, and in Sec.~V we calculate the achievable secret key rates. Finally in Sec.~VI we present our conclusions.

\section{A Practical CKA Implemetation}\label{sec:bvgiheeijorsdf}\noindent
The quantum CKA protocol we consider in this paper requires $N$-party GHZ states of the form of Eq.~(\ref{eq:mngero}). However, such states are difficult to create naturally due to the lack of $N$-party interaction Hamiltonians. Instead one can create a number of bi-partite Bell states, and entangle them to obtain an $N$-party GHZ state \cite{Fedrizzi_2007}. This reduces the creation of GHZ states to the creation of Bell states and the ways to entangle them further.

\begin{figure}[t!]
\centering
\includegraphics[width=0.5\columnwidth]{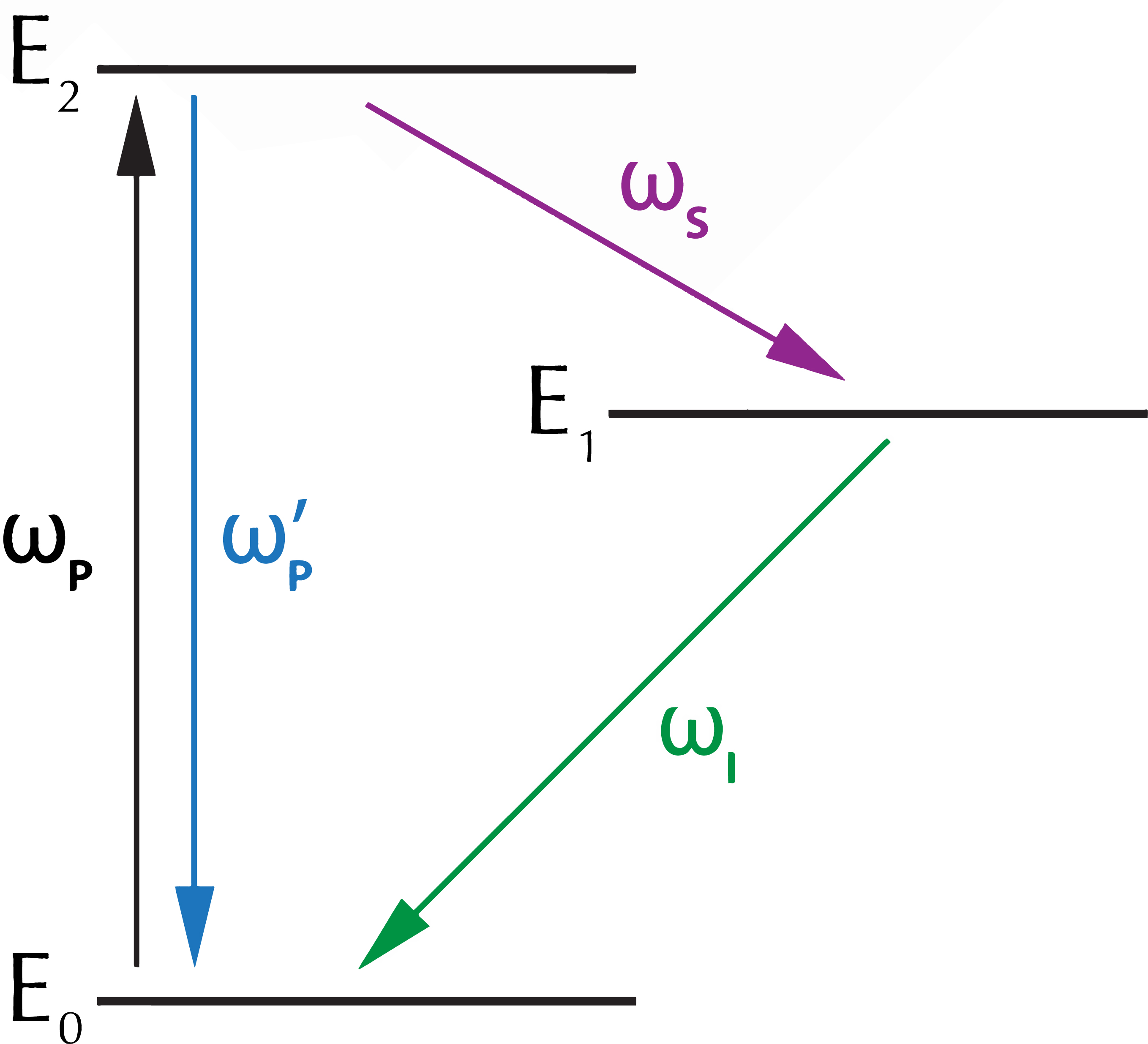}
\caption{Transition energy levels in a nonlinear material responsible for parametric downconversion. Photon frequencies for the pump, signal and idler are represented by $\omega_P$, $\omega_s$ and $\omega_i$, respectively. The energy levels $E_0$ represents the ground state, where $E_1$ and $E_2$ represent excited states.}
\label{PDClevel}
\end{figure}

Currently, two-photon Bell states are almost exclusively created using a process called parametric downconversion (PDC). The main mechanism of PDC is a cascade in nonlinear materials, as shown in Fig.~\ref{PDClevel}. The most commonly used materials are KDP ($\mathrm{KD_2PO_4}$) and BBO ($\beta$-$\mathrm{BaB_2O_4}$) \cite{gerry2005introductory}. A pump laser with frequency $\omega_P$ excites the material to an excitation level $E_2$. With high probability the material decays back directly to $E_0$, but there is a small probability of decaying back down to level $E_0$ via level $E_1$. In this case, the material emits two photons with frequencies $\omega_s$ (the ``signal'' photon) and $\omega_i$ (the ``idler'' photon). Conservation of energy requires that $\omega_s+\omega_i = \omega_P$, and momentum conservation requires $\mathbf{k}_s + \mathbf{k}_i = \mathbf{k}_P$, where $\mathbf{k}_j$ denotes the wave vector of mode $j$. These are the phase matching conditions.

In type-II PDC, the nonlinear crystals are arranged such that the signal and idler photons have opposite polarisation. The interaction Hamiltonian describing this process can be written as 
\begin{align}
H_{\rm PDC} = \xi (\hat{a}^\dag_{H,s} \hat{a}^\dag_{V,i} -\hat{a}^\dag_{V,s} \hat{a}^\dag_{H,i}) + \text{H.c.,}
\label{eq2.34}
\end{align}
where $\xi$ is the coupling strength of the downconversion process, $\hat{a}^\dagger_j$ is the creation operator for a photon in mode $j$, and H.c.\ stands for Hermitian conjugate. The resulting state in the signal and idler modes then becomes 
\begin{align}
\ket{\psi} &= {\rm e}^{-iH_{\rm PDC}t/\hbar}\ket{0} \equiv \sqrt{1-\lambda^2}\sum_{k=0}^\infty \lambda^k \ket{\Phi_k},
\label{eq2.35}
\end{align}
where $\lambda\in[0,1)$ depends on the strength of the pump laser and the thickness and nonlinearity of the material, and $t$ is the duration of the pump pulse. The states $\ket{\Phi_k}$ are given by \cite{Kok_2000}
\begin{align}
\ket{\Phi_k} = \frac{1}{\sqrt{k+1}}\sum_{m=0}^k (-1)^m \ket[si]{m,k-m;k-m,m},
\label{eq2.36}
\end{align}
with $\ket[si]{m,k-m;k-m,m}$ the state of $m$ photons in the horizontal signal and the vertical idler modes, and $k-m$ photons in the vertical signal  and the horizontal idler modes. As a result, the PDC process does not produce pure two-photon Bell pairs, but rather a superposition of different numbers of photon pairs. Creating no photon pairs at all ($k=0$) is most likely, and more photon pairs become increasingly unlikely for typical values of $\lambda \sim 10^{-2}$.

\begin{figure}[t!]
\centering
\includegraphics[width=0.35\textwidth]{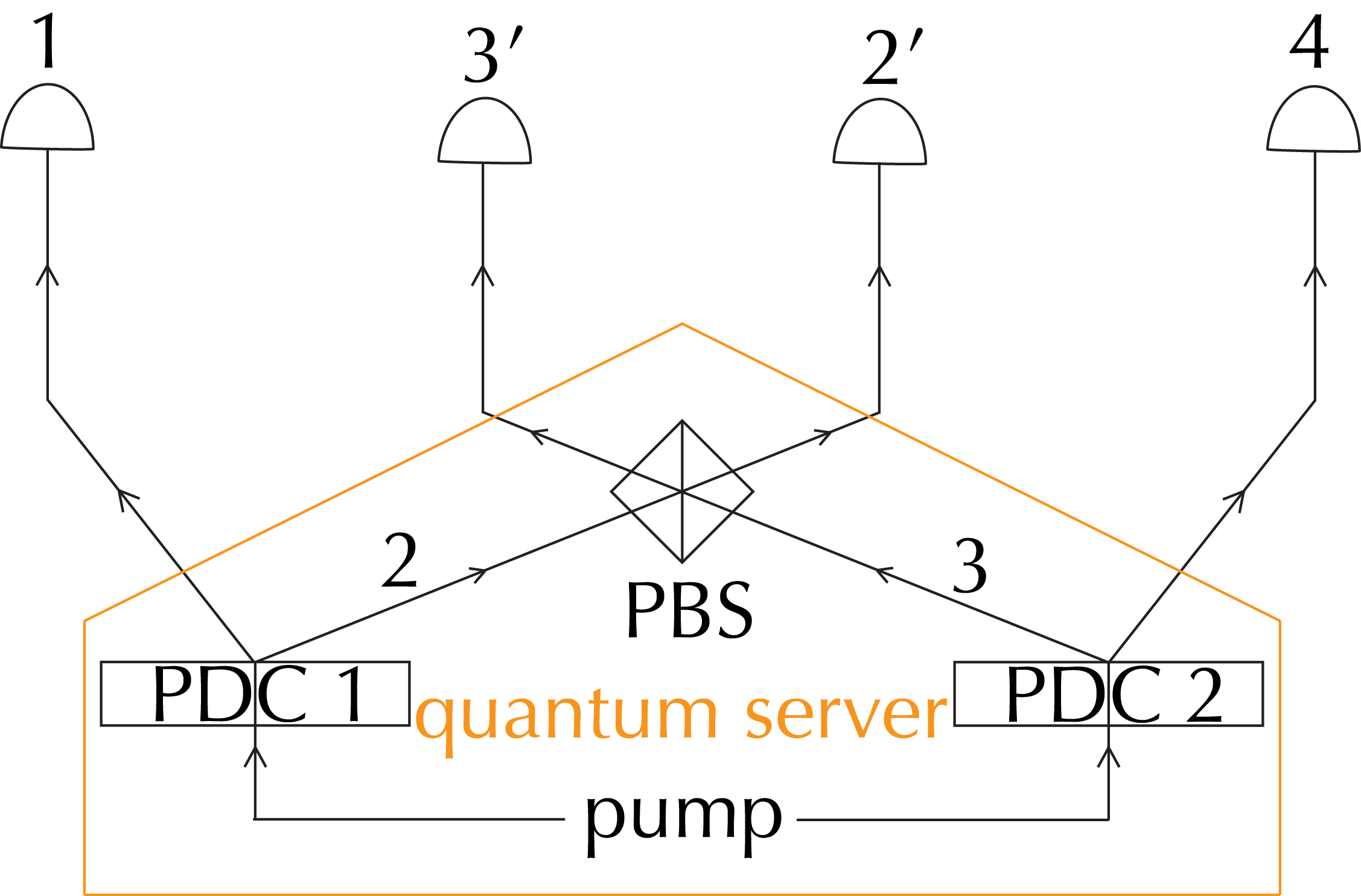}
\caption{Schematic diagram of a CKA protocol following \cite{proietti2020experimental}, an experiment to create 4 parties entanglement via sharing a four-GHZ state. Entangled photon is created in the server represented by the orange box. PDCs are parametric downconverters. The box labelled PBS at the middle is the polarisation beam splitter. Photons are then sent to each parties, labelling 1 to 4, the primed modes are photon modes after exiting the PBS. Post-measurement is performed using bucket detectors.}
\label{setup}
\end{figure}

To create an $N$-party GHZ state for the CKA protocol, we can use a chain of PDCs and mix the signal modes of two adjacent PDCs onto a polarising beam splitter (PBS), which transmits horizontally polarised light, and reflects vertically polarised light. The PBS can be considered part of the central server, as shown in Fig.~\ref{setup}. For illustrative purposes we consider the case of four parties in the CKA protocol ($N=4$), which is the version implemented in the experiment performed by Proietti \emph{et al.}\ \cite{proietti2020experimental}. 

We are interested in the events that give pairs of entangled Bell states $\ket{\Phi_1}$ in the two PDCs. However, with comparable probability, one of the PDCs will create two pairs, while the other PDC does not create any pairs. Up to $O(\lambda^3)$, the output state of the four modes is given by
\begin{align}
\ket{PDC_1}\ket{PDC_2} \propto \;
 & \ket[12]{\Phi_0}\ket[34]{\Phi_0}  \cr
 &  + \lambda\ket[12]{\Phi_1}\ket[34]{\Phi_0} + \lambda\ket[12]{\Phi_0}\ket[34]{\Phi_1}  \cr
 &  + \lambda^2\ket[12]{\Phi_0}\ket[34]{\Phi_2} + \lambda^2 \ket[12]{\Phi_2}\ket[34]{\Phi_0}  \cr
 &  + \lambda^2\ket[12]{\Phi_1}\ket[34]{\Phi_1} + O(\lambda^3)\, .
\label{eq3.1}
\end{align}
The subscripts 1, 2, 3, and 4 refer to the spatial modes in Fig.~\ref{setup}.  Eq.~(\ref{eq3.1}) is a very good approximation when $\lambda\ll1$. 

Each party must receive a photon in the CKA protocol, which means that the only term in Eq.~(\ref{eq3.1}) that is of use to us is $\ket[12]{\Phi_1}\ket[34]{\Phi_1}$. In the CKA protocol, we can post-select on measuring photons in all four spatial modes, which exactly singles out this term. However, when we do that, we no longer have a freely propagating state, since photodetectors are destructive. After measuring a photon, the photon has disappeared as energy in the detector. Therefore, when we refer to the creation of a GHZ state it is important to remember that we mean a \emph{post-selected} state: given that each party receives exactly one photon, the state prior to detection was the GHZ state. From now on, we will understand the creation of GHZ states in this way.

Next, modes 2 and 3 are sent into in the PBS, and transformed into modes $2^\prime$ and $3^\prime$. The term $\ket[12]{\Phi_1}\ket[34]{\Phi_1}$ is then transformed into the state on modes 1, $2^\prime$, $3^\prime$, and 4 as
\begin{align}\nonumber
 \frac{1}{\sqrt{3}}\ket{H;0;HV;V} & + \frac{1}{2}\ket{H;V;V;H} \cr & + \frac{1}{2}\ket{V;H;H;V} + \frac{1}{\sqrt{3}}\ket{V;HV;0;H}\, ,
\end{align}
where the PBS has caused the two photons from modes 2 and 3 to sometimes both go to $2^\prime$ or to $3^\prime$. Further post-selection on finding exactly one photon in mode $2^\prime$ and in mode $3^\prime$ then allows us to infer that the impinging optical field was in the state 
\begin{align}\nonumber
 \frac{1}{2}\ket{H;V;V;H}+\frac{1}{2}\ket{V;H;H;V}\, .
\end{align}
This state is not normalised, reflecting the reduced probability of finding the four photons arriving at four different parties. Note that the communicating parties still have access to the entanglement in this post-selected state since they can freely choose the polarisation basis in which to measure the incoming photon. GHZ states with a larger number of parties can be constructed from chaining more PDCs and mixing modes on a PBS. The success probability of this method reduces exponentially in $N$.


Since the protocols presented here operate in a post-measurement fashion, good photodetectors are required. Photodetectors can be categorised into two main groups, bucket detectors and number-resolving detectors. The bucket detector is able to tell if there is at least one photon presented but is unable to tell how many photons are there. The number-resolving detector, however, is able to tell how many photons have been detected. Good detectors have low dead time, low dark count rates, low time jitter, and low photon loss $\eta_d$. There are many ways to implement bucket detectors. The most common devices are photomultiplier tubes and avalanche photodiodes \cite{kok2010introduction}. Number-resolving detectors can be constructed using a variety of physical implementations, including superconducting transition-edge sensors, superconducting nanowire single-photon detectors and single-photon detectors based on quantum dots and semiconductor defects \cite{Hadfield}. 


\section{Photon loss \& Error Correction}\label{sec:vjgheru9iowjs}\noindent
In this section we consider in detail the effect of photon loss on the secret key rate of the quantum CKA protocol. We then review the parity and redundant encoding for qubits that can be used to mitigate these photon losses.

\subsection{Photon Loss}\noindent
Losing a photon in a long transmission line is common in fibre optics. Party $A$ in our protocol can be assumed to have no transmission loss because it is very close to the server. For party $B_i$, loss depends on the distance that photons have to travel in the fibre of length $l_i$. The constant of loss is the \emph{attenuation length}, $l_0$, relating to the loss probability, $\eta_i$, by
\begin{align}
 \eta_i = 1-e^{-l_i/l_0}\, .
\label{eq1.1}
\end{align}
Typical optical fibers have attenuation rate of about 0.1 dB/km or less \cite{Weik2001}. Photon loss also occurs in each party when the photons enter photodetectors. We assume that every party has identical detectors with loss probability $\eta_d$. The transmission and detection losses are independent, and the total loss probability is given by
\begin{align}
 \eta_{\text{tot},i} = \eta_i \cdot \eta_d\, .
\label{eq1.2}
\end{align}
In real experiments we generally do not know where the loss occurs, so $\eta_{\text{tot},i}$ is the appropriate parameter to consider. In our analysis, we will assume that every party $B_i$ has the same loss probability, which we denote by $\eta$. The cases with different loss probability in each party, i.e., $\eta_i \neq \eta_j$ when $i\neq j$, can be straightforwardly generalised but is algebraically more involved. 

In order to preserve the entanglement for a successful protocol, we cannot afford to lose any photons at all. Hence, the success probability of the protocol is equal to the probability of every party receiving its photon:
\begin{align}
 p(\text{success,n}) = p(\text{no loss at all}) = (1-\eta)^n\, .
\label{eq3.10}
\end{align}
This is the transmission probability of the existing CKA protocol, i.e., $p(\text{success,n}) = p(\text{transmit})$. This makes photon loss a catastrophic failure for the protocol, and as a result it will be quickly outperformed by BB84 if no measures are taken to deal with photon loss. 

\subsection{Error Correction}\label{sec:nehurijwp}\noindent
To salvage the quantum CKA protocol, we have to mitigate photon loss. This can be achieved using  error correction. The encoded logical qubits, denoted with the subscript $\ket[L]{\cdot}$, then consist of states of many physical qubits. We will be using two types of encoding for the error correction process, namely parity encoding and redundant encoding. We will introduce both protocols in a general computational basis representation $\{0,1\}$, which maps directly onto the polarisation  representation $\{H,V\}$. 

\paragraph*{Parity encoding}
The logical qubits of parity encoding are defined as
\begin{align}
\ket[L]{0}=\ket{0}^{(m)} &\equiv \frac{1}{\sqrt{2}}(\ket{+}^{\otimes m}+\ket{-}^{\otimes m}),\cr
\ket[L]{1}=\ket{1}^{(m)} &\equiv \frac{1}{\sqrt{2}}(\ket{+}^{\otimes m}-\ket{-}^{\otimes m})\, ,
\label{eq2.50}
\end{align}
where $\ket{\pm} = (\ket{0}\pm\ket{1})/\sqrt{2}$. The parity encoded qubits can also be defined recursively via
\begin{align}
\ket{0}^{(m)} &= \frac{1}{\sqrt{2}}\left(\ket{0}\ket{0}^{\otimes m-1}+\ket{1}\ket{1}^{\otimes m-1}\right),\cr
\ket{1}^{(m)} &= \frac{1}{\sqrt{2}}\left(\ket{1}\ket{0}^{\otimes m-1}+\ket{0}\ket{1}^{\otimes m-1}\right).
\label{eq2.51}
\end{align}
The recursive definition is useful when considering how a parity encoded qubit could protect the state when there is photon loss. Consider modelling photon loss as a measurement in computational basis, \{0,1\}, without knowing the result. Using this loss model, we can see that the outcome 0 gives $\ket{0}^{(m-1)}$ and the outcome 1 gives $\ket{1}^{(m-1)}$, which are still in the form of \eqref{eq2.50}, with $m$ reduced by 1. Similarly for $\ket{1}^{(m)}$, the measurement outcome 0 gives $\ket{1}^{(m-1)}$ and the outcome 1 gives $\ket{0}^{(m-1)}$. Hence, one photon loss either leaves the logical qubit the same or flip the qubit once. Since we do not know which, the qubit state is a mixture of the two.

The encoded states in Eq.~(\ref{eq2.50}) are highly entangled, and we need to use entangling gates such as the \textsf{CNOT} to create these states. However, such gates are problematic in linear optics. Instead, we can use the so-called fusion gates to create these states \cite{kok2010introduction}. Type-I ($\mathcal{F}_{I}$) and type-II ($\mathcal{F}^\prime_{II}$) fusion gates \cite{Browne_2005} are used to create optical \textsf{CNOT} gates for parity logical qubits. 

\paragraph*{Redundant encoding}
Parity encoding can protect qubits when photon loss occurs. However, since the photon is lost, we do not know the measurement result, hence, we have no way of knowing whether the parity qubit has been flipped or not. Hence, another level of encoding is required. This is provided by the redundant encoding. In this quantum error correction code, the logical qubits are given by
\begin{align}
\ket[L]{0} &=\ket{0}^{(m,q)}\equiv \ket[1]{0}^{(m)} \otimes \ket[2]{0}^{(m)} \otimes \dots \ket[q]{0}^{(m)}, \cr
\ket[L]{1} &=\ket{1}^{(m,q)}\equiv \ket[1]{1}^{(m)} \otimes \ket[2]{1}^{(m)} \otimes \dots \ket[q]{1}^{(m)},
\label{eq2.52}
\end{align}
where $\ket[i]{0}^{(m)}$ and $\ket[i]{1}^{(m)}$ with $i =1,2,\dots,q $ are defined in Eq.~\eqref{eq2.50}. We have included the parity encoding in this description. The purely redundant encoding is retrieved for $m=1$. The error correction of a lost photon can now be achieved by projecting the state of the lost photon onto $\ket[L]{0}$ or $\ket[L]{1}$, thus providing the missing information needed to reconstruct the pure quantum state.


\section{Loss Tolerant CKA Protocol} \label{sec:fhieru9iowj}
\noindent
In this section we will present a protocol to protect the GHZ state from loss. The protocol is based on parity encoding and redundant encoding with error correction facility in each party. 

\subsection{Experimental Setup with Encoding} \label{sec4.1}
\begin{figure}[t!]
\centering
\includegraphics[width=0.3\textwidth]{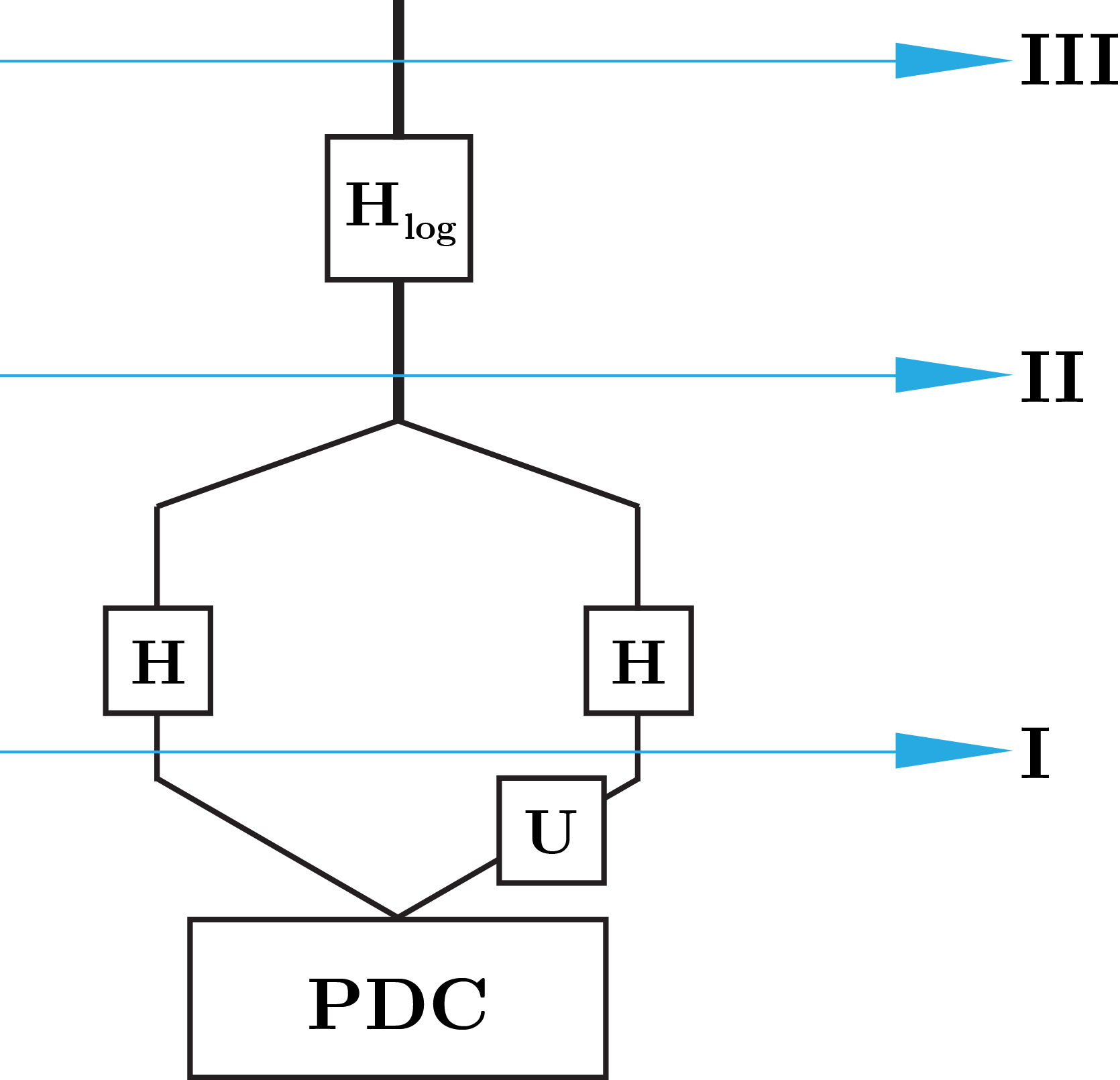}
\caption{Schematic diagram of parity encoded qubit generation. The PDC produces a Bell state and it is transformed  into the required type of Bell state using a unitary operator \textsf{U} (Stage I). Hadamard gates \textsf{H} are applied to convert the state into parity encoding (Stage II). $\textsf{H}_{\text{log}}$ is then applied to transform a logical zero into logical plus state.}
\label{circ_one}
\end{figure}

\noindent
To implement our protocol we must create a parity encoded Bell state, following the stages shown in Fig.~\ref{circ_one}. The first stage creates the state $\ket{\psi_I}$, consisting of a Bell state generated using PDC and a unitary operator \textsf{U}. The unitary operator converts the standard PDC output Bell state to the required Bell state. Although the protocol works for any Bell state, we will demonstrate the protocol using a $\ket{\Phi^+}$ state, given by
\begin{align}
\ket{\psi_I} = \ket{\Phi^+} = \frac{1}{\sqrt{2}}(\ket{00}+\ket{11})\, .
\label{eq4.1}
\end{align}
The state in \eqref{eq4.1} is then encoded into a parity encoding qubit using Hadamard gates \cite{Cerf_1998}, resulting in the Stage II state given by
\begin{align}
\ket{\psi_{II}} = \frac{1}{\sqrt{2}}(\ket{++}+\ket{--}) =\ket{0}^{(2)},
\label{eq4.2}
\end{align}
which is a parity encoded qubit. For Stage III, we want to turn $\ket{0}^{(2)}$ into $\ket{+}^{(2)}$ using a Hadamard on the logical qubit $\textsf{H}_{\text{log}}$, producing
\begin{align}
\ket{\psi_{III}} = \ket{+}^{(2)}= \frac{1}{\sqrt{2}}(\ket{0}^{(2)}+\ket{1}^{(2)}).
\label{eq4.3}
\end{align}
This is the basic building block for creating larger encoded states.

\begin{figure}[t!]
\centering
\includegraphics[width=0.5\textwidth]{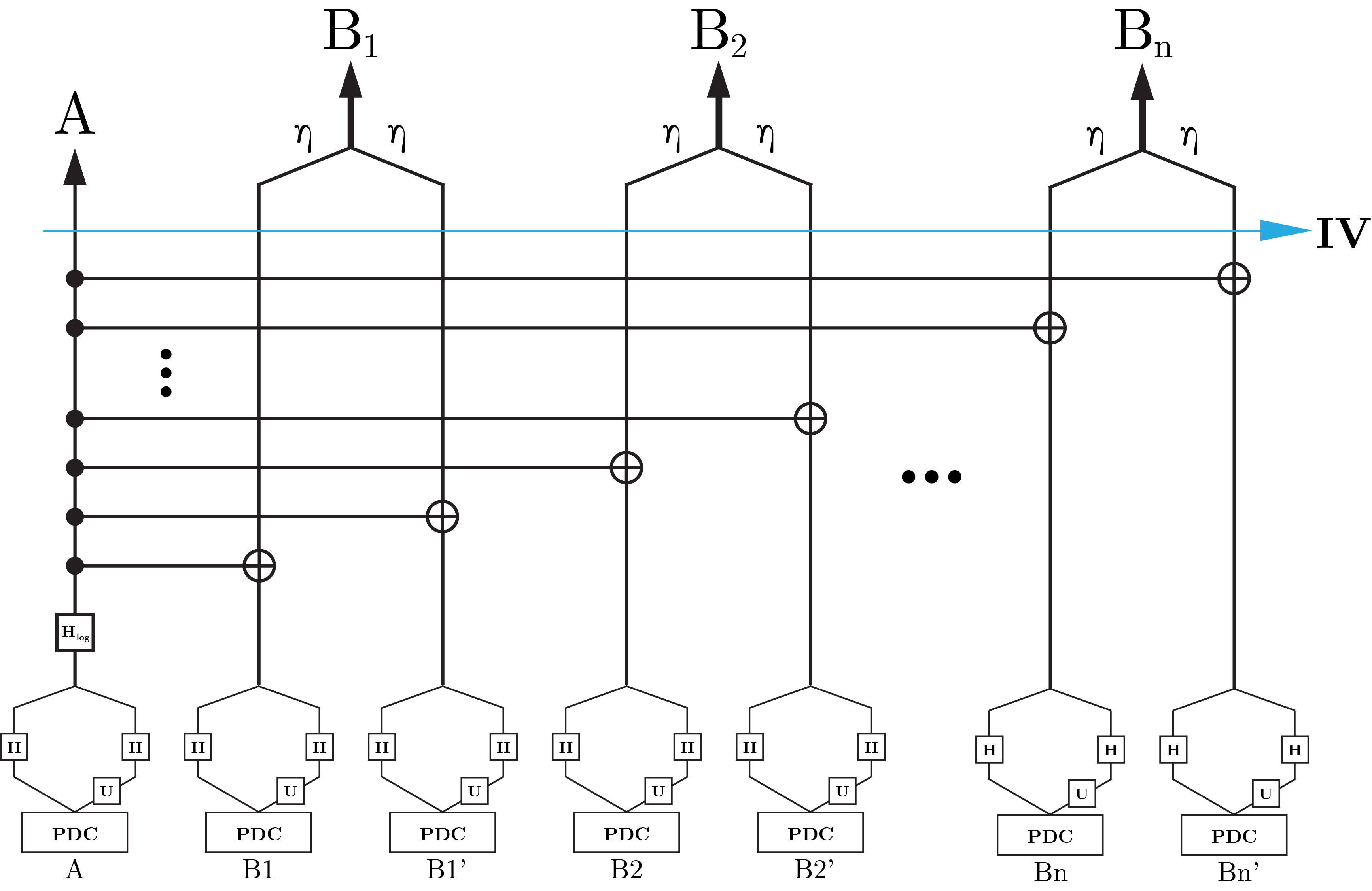}
\caption{Experimental setup for $N$ parties redundant encoding. Party $A$ is the server distributing entanglement to parties $B_i$ when $i = 1,2,\dots,n$ with loss probability $\eta$ per photon}
\label{circn_all}
\end{figure}

To perform redundant encoding, we connect the circuit in Fig.~\ref{circ_one} together with parity encoded circuits using \textsf{CNOT} gates. The entire circuit is shown in Fig.~\ref{circn_all}. The left most subcircuit is identical to Fig.~\ref{circ_one}, while the remaining subcircuits are equal to Fig.~\ref{circ_one} without Stage III.  This is the blueprint for our loss-resilient encoded protocol.

The state in Stage IV, after the operation of $2n$ \textsf{CNOT} gates, is given by
\begin{align}
\ket{\psi_{IV}} =\frac{1}{\sqrt{2}}(\ket[L]{0}^{\otimes 2n+1}+\ket[L]{1}^{\otimes 2n+1}),
\label{eq4.4}
\end{align}where each $\ket[L]{0}$ and $\ket[L]{1}$ are logical parity encoded qubits of each party $A, B_1, B_2, \dots, B_n$. Two logical qubits are sent to each party $B_i$, while only one logical qubit is sent to party $A$. If there is no photon loss, each party will receive their photons from the state in Eq.~\eqref{eq4.4}.

\subsection{Error Correction} \label{sec4.2}
\begin{figure}[t!]
\centering
\includegraphics[width=0.4\textwidth]{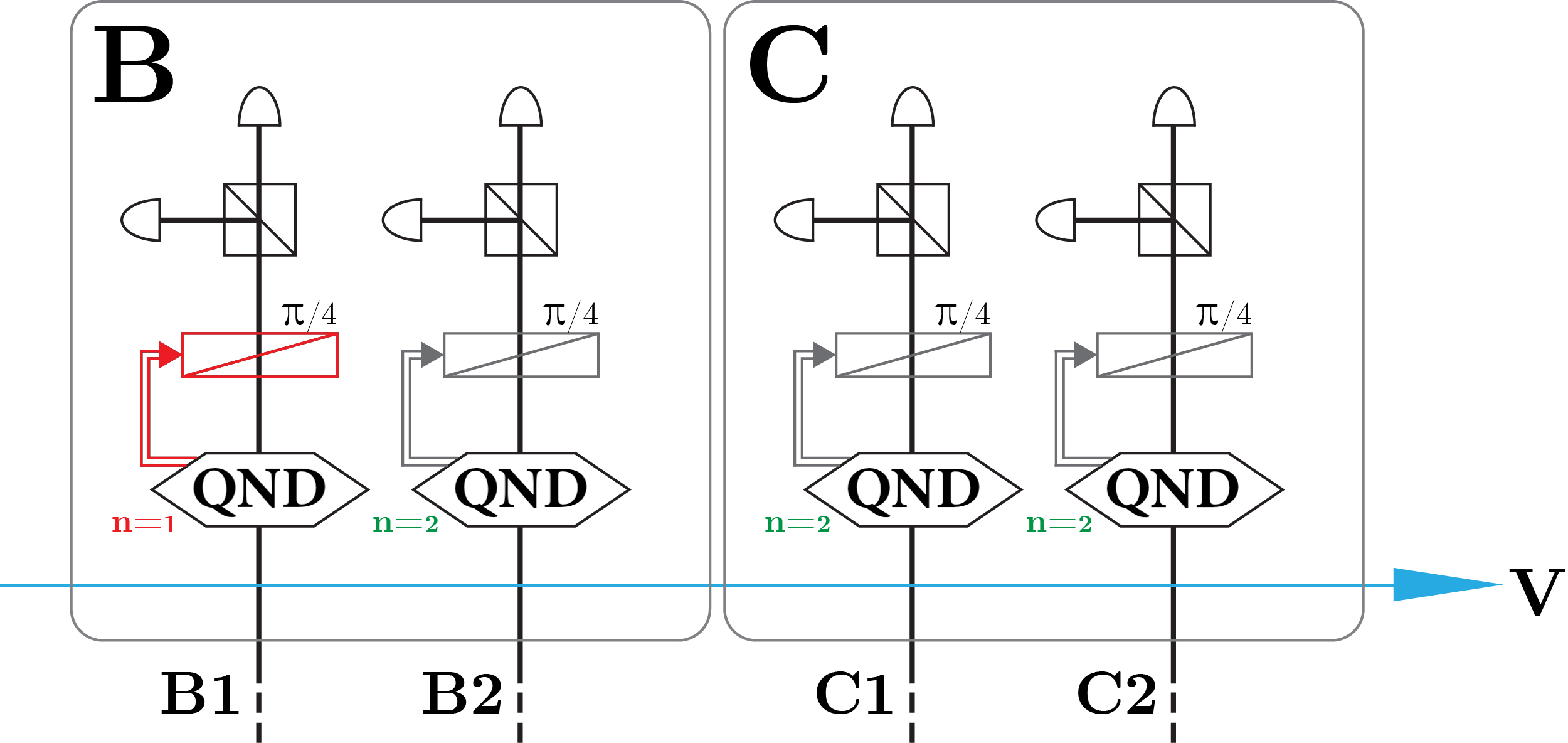}
\caption{The error correction facility in each party. The setup in party $B$ represents the situation where one photon is lost. The setup in party $C$ represents the situation with no photon loss. The QNDs are non-demolition photon number detectors, the output of which triggers a phase shifter using classical channels. The detectors at the top are regular photodetectors.}
\label{correct1}
\end{figure}

\noindent
After receiving the logical qubits, each party has its own error correction facility. The facility allows each party to correct their qubits if there is photon loss, and retain entanglement between all parties. The error correction facility is given in Fig.~\ref{correct1}, with a scenario with photon loss (left) and a scenario with no photon loss (right). The facility for each mode consists of a quantum-nondemolition detector (QND) \cite{Guerlin_2007,inbook,birnbaum2005photon,Wilk488,Kok_2002,PhysRevLett.57.2473,Nogues1999}, a $\pi/4$ polarisation rotation and a polarisation photon detector. A QND detector can measure the number of photons in a mode without destroying the photon. If there is photon loss, the classical channel linking QND to polarisation rotation will send signal to rotate the polarisation rotation, otherwise, the polarisation rotation is left idle. Finally, the photons are measured by a polarisation photon detector, which can be implemented using a PBS and two detectors at the top of each mode. The measurement is originally in computational basis $\{0,1\}$, however, with the polarisation rotation triggered, the measurement changes into the diagonal basis $\{+,-\}$. The situation when there is one photon loss (in mode $B1$) is demonstrated by party $B$ in Fig.~\ref{correct1}, where a classical signal (in red) switches a phase shift inducing a polarisation rotation. When there is no photon loss (demonstrated by party $C$ in Fig.~\ref{correct1}) there is no classical signal from the QND detectors (in grey). 

\begin{figure}[t!]
\centering
\includegraphics[width=0.33\textwidth]{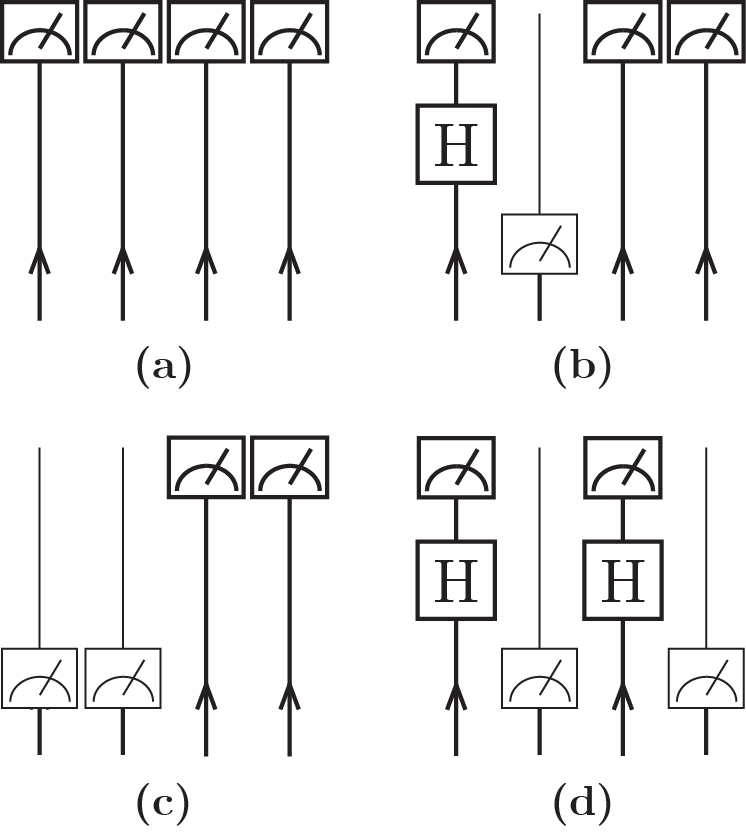}
\caption{Summary of error correction step for four different cases in one party receiving two modes (four photons). The two upper cases, (a) and (b), result in successful protocol. The two lower cases, (c) and (d), result in protocol failure.}
\label{error_sum}
\end{figure}

Next, we describe the error correction process for four scenarios in each party, as shown in Fig.~\ref{error_sum}. This includes (a) no photon loss, (b) one loss in one mode, (c) two loss in one mode and (d) one loss in each mode. Modes with photon loss are represented by thinner lines terminated by measurement. Sample states below are given for three-party entanglement, $A,B$ and $C$, where party $A$ is not shown in Fig.~\ref{correct1}. One logical qubit is sent to party $A$, while two logical qubits are sent to parties $B$ and $C$ each.

\subsubsection*{(a) No Photon Loss}\noindent
Since there is no loss, the error correction facility is not activated. This is equivalent to the situation for party C in Fig.~\ref{correct1}. Both QNDs detect that both modes $C1$ and $C2$ have $n=2$ photons, i.e., no photon loss. The result $n=2$ leaves the polarisation rotations inactive, hence, the measurements are still in the computational basis. The state after post-measurement is inferred to be an encoded GHZ state, given by
\begin{align}
\ket[\text{(a)}]{\psi_{V}} =\frac{1}{\sqrt{2}}(\ket[L]{0}^{\otimes 5}+\ket[L]{1}^{\otimes 5}),
\label{eq4.5}
\end{align}where the five modes are $A$, $B_1$, $B_2$, $C_1$ and $C_2$.

\subsubsection*{(b) One Photon Loss in One Mode}\noindent
In this scenario, the error correction facility will be activated for the mode with photon loss. Consider party $B$ in Fig.~\ref{correct1}. The QNDs detect that mode $B1$ has $n=1$ photon and mode $B2$ has $n=2$ photons. The classical signal generated by the QND for $n=1$ triggers the polarisation rotation. Hence, measurement in mode $B1$ is changed into diagonal basis. The measurement still remains in computational basis for mode $B2$.

We will show that, if zero or one photon is lost in a party, the entanglement sharing can be recovered. However, if more than one photon is lost in a party, the protocol fails. The state after measurement in diagonal basis in mode $B1$ is given by
\begin{align}
\ket{\psi_{V,B1\text{gone}}} &= \begin{cases}
      \text{outcome } + :\frac{1}{\sqrt{2}}(\ket[L]{0}^{\otimes 4}+\ket[L]{1}^{\otimes 4})\\
      \text{outcome } -:\pm\frac{1}{\sqrt{2}}(\ket[L]{0}^{\otimes 4}-\ket[L]{1}^{\otimes 4}),
    \end{cases}
\label{eq4.6}
\end{align}where the four modes are $A$, $B_2$, $C_1$ and $C_2$. It is still an encoded GHZ state between three parties up to phase flips. The plus and minus signs in $\ket{\psi_{V,B1\text{gone}}}$ do not affect the probability in a measurement outcomes. Note also that Eq.~\eqref{eq4.6} needs to be post-selected by measuring all the remaining modes in computational basis, hence, it will be an inferred state with no real propagating photons.

We have shown that the error correction works for one photon loss in a party. Consider further when there is another photon loss in party $C$. Without loss of generality, let the loss occur in mode $C1$. In this case, QND in $C1$ detects $n=1$ photon and QND in $C2$ detects $n=2$ photons. Similar error correction is performed in party $C$, giving the inferred state after post-measurement to be
\begin{align}
\ket{\psi_{V,B1\text{gone},C1\text{gone}}} = \pm\frac{1}{\sqrt{2}}(\ket[L]{0}^{\otimes 3}\pm\ket[L]{1}^{\otimes 3}),
\label{eq4.7}
\end{align}where the three modes are $A$, $B_2$ and $C_2$. This is also an encoded GHZ state between three parties. Again, the plus and minus signs do not affect measurement probability. Hence, for $N$ parties with one photon loss in each party, the error correction process can be performed accordingly and we can recover all parties entanglement.

\subsubsection*{(c) Two Photon Loss in One Mode}\noindent
Losing two photons in one mode is equivalent to losing an entire logical qubit. Since we are sending an encoded GHZ state, losing one of the encoded states results in losing all the entanglement. Hence, the resulting state before entering each party (Stage V) is given by
\begin{align}
\ket[L]{0}^{\otimes 4} \text{ or } \ket[L]{1}^{\otimes 4},
\label{eq4.8}
\end{align}where the four modes are $A$, $B_2$, $C_1$ and $C_2$. The state is not entangled, which means we have already lost the entanglement between all parties. Hence, the error correction process is unable to recover the entanglement when at least one of the QNDs detects $n=0$ photon.

\subsubsection*{(d) One Photon Loss in Each Mode}\noindent
This scenario happens when both QNDs of party $B$ detect $n=1$ photon. It is similar to an extended consideration in (b), where we have considered photon loss in modes $B_1$ and $C_1$. Here, we experience photon loss in modes $B_1$ and $B_2$, instead. Hence, the error correction process is similar to what was done in (b), giving the resulting state as \eqref{eq4.7} but with the three modes being $A,C_1$ and $C_2$.

There is no mode belonging to party $B$ left in the encoded GHZ state. Hence, party $B$ has been excluded from the system. It is better than completely losing the entire entanglement because other parties are still entangled, however, we have fail to retain entanglement between all parties.


\section{Achievable Secret Key Rates}\label{sec:bfdhrjdfnkvc}
\noindent
We now have a complete description of how the encoded CKA protocol works, including its limitations. This section compares the performance of our encoded protocol and the existing non-encoded protocol. We will perform quantitative analysis on entanglement creation rate and entanglement transmission rate for both protocols. The total secret key rate, used to determine protocol performance, is the product of these two rates. We assume for simplicity that each party $B_i$ experiences the same total loss $\eta$.

\subsubsection*{Entanglement transmission rate}
\begin{figure}[t!]
\centering
\includegraphics[width=\columnwidth]{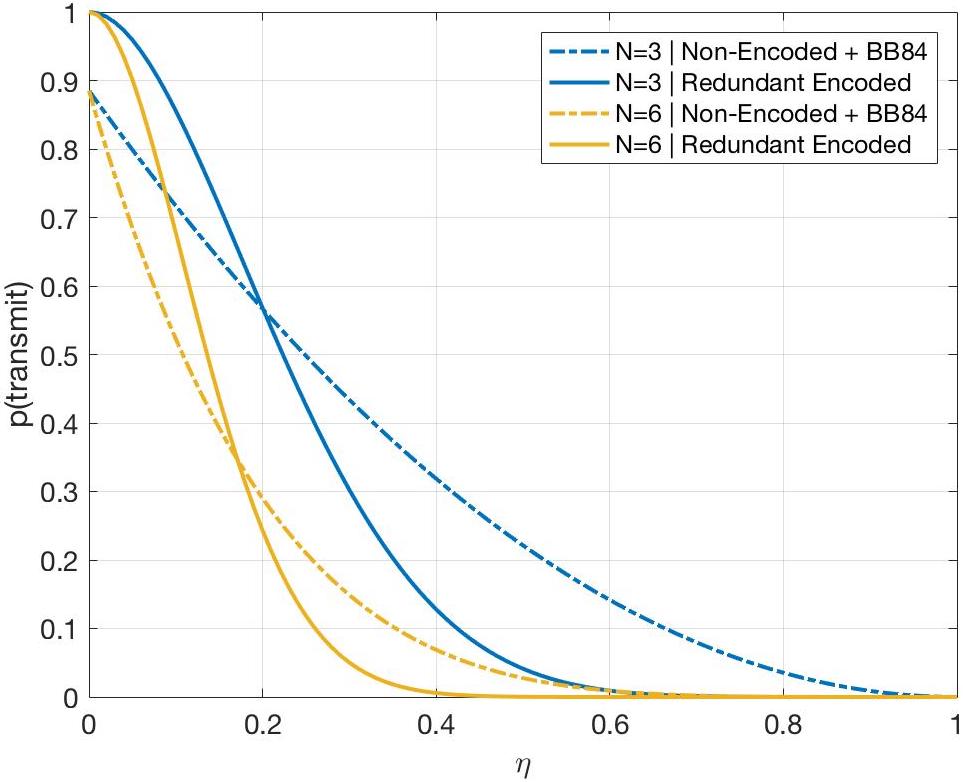}
\caption{Probability of successful entanglement transmission as a function of loss probability $\eta$ in three ($N=3$, blue) and six ($N=6$, yellow) parties entanglement distribution using non-encoded with $N$-BB84 protocol (dashed line) and 2-photon redundant encoded protocol (solid line).}
\label{p-eta}
\end{figure}

\noindent
The entanglement transmission rate is proportional to the entanglement transmission probability. The success probability of the existing non-encoded CKA protocol is given by \cite{proietti2020experimental}
\begin{align}
 p(\text{success,n})_\text{non-enc} = (1-\eta)^n \times \frac{(L-2m)(1-h(p))}{L}\, ,
 \label{eq4.9}
\end{align}
where $h(p)$ is the single bit entropy function 
\begin{align}
 h(p) = - p \log_2 p - (1-p) \log_2 (1-p)\, .
\end{align}
Next, we will find the entanglement transmission probability for our protocol. As discussed in the last section, error correction facilities allow each party to cope with one photon loss. The success probability for one party is therefore given by
\begin{align}
p(\text{success,1})_\text{enc} &= p(\text{no loss}) + p(\text{1 photon lost})\cr
&= (1-\eta)^4 + \begin{pmatrix}
4\cr
1
\end{pmatrix}(1-\eta)^3 \eta\cr
&=1-6\eta^2+8\eta^3-3\eta^4.
\label{eq4.10}
\end{align}
For $N$ parties, the success probability is the product of every party's success probability, given by
\begin{align}
p(\text{success,n})_\text{enc} &= p(\text{success,1})^n\cr
&=(1-6\eta^2+8\eta^3-3\eta^4)^n.
\label{eq4.11}
\end{align}
To compare the entanglement transmission rate of both protocols, \eqref{eq4.9} and \eqref{eq4.11} are plotted in Fig.~\ref{p-eta} for situations with $N=3$ and $N=6$ parties. As expected, the lower the number of parties, the higher transmission probability, hence, the higher transmission rate. For the non-encoded protocol (dashed lines), we can see that the probability is less than one even if with $\eta = 0$, because BB84 protocol will always reduced the key rate by trading it with security.

\subsubsection*{Entanglement creation rate}\noindent
Next, we study the entanglement creation rate by comparing the probability of creating entangled states for both protocols. Since each device is assumed to be independent, the successful creation probability comes from the product of each element's success probability. All optical devices required to construct the protocol are considered. The total number of parties is $N = n+1$, where $n$ is the number of parties $B_i$ with loss and another party is the party $A$ without trransmission loss.

First, we determine how many resources are required to create entanglement in a non-encoded protocol. The $N$-party version of Fig.~\ref{setup} uses one PDC per two parties that reduce the success probability of the protocol and need to be taken into account. Hence, for $n+1$ parties, the minimum requirement is $\lceil (n+1)/2\rceil$ PDCs. There are also PBSs, which are required one fewer than the number of PDCs, hence, $\lfloor n/2\rfloor$. Each transmission line is subjected to loss probability of $\eta$. 

Next, consider the resource needed to create entanglement in the presented protocol. As shown in Fig.~\ref{circn_all}, the protocol uses two PDCs per party $B_i$ with loss, hence, $2n+1$ PDCs overall. There are also unitary gates, $2n+1$ \textsf{U} gates, $2(2n+1)$ \textsf{H} gates, 1 $\textsf{H}_\text{log}$ gate and $2n$ \textsf{CNOT} gates for parity encoded qubits. Each transmission line is subjected to a loss probability of $\eta$.
Table 3 summarises probability of creating entanglement for both protocols in the server.

\begin{table}[t!]
\caption{\label{tab:poretjngkf}Probability of creating entanglement for existing and presented protocols.}
\begin{ruledtabular}
\begin{tabular}{lcc}
 Optical Elements &Existing (Fig.~\ref{setup})&Presented(Fig.~\ref{circn_all}) \\ 
 {PDC}s & $\left(\lambda^2(1-\lambda^2)\right)^{\lceil (n+1)/2\rceil}$ & $\left(\lambda^2(1-\lambda^2)\right)^{2n+1}$ \\ 
 {PBS}s & $2^{-\lfloor n/2\rfloor}$ &-- \\ 
 Quantum Gates & -- & $4^{-2n} $ \\
\end{tabular}
\end{ruledtabular}
\end{table}

To see how to obtain each row and column in Table~\ref{tab:poretjngkf}, consider the arguments below. The first row comes from the probability of getting a post-selected Bell state from PDC. The state of the PDC is given by
\begin{align}
\ket{\psi_\text{PDC}} = \sqrt{1-\lambda^2}\sum_{n=0}^\infty \lambda^n \ket{\Phi_n},
\label{eq4.12}
\end{align}
where the required Bell state is $\ket{\Phi_1}$,
\begin{align}
\ket{\Phi_1} =\frac{1}{\sqrt{2}}(\ket[S]{V}\ket[I]{H}-\ket[S]{H}\ket[I]{V}).
\label{eq4.13}
\end{align}
The probability of producing this state is 
\begin{align}
p(\text{Bell}) =\lambda^2(1-\lambda^2).
\label{eq4.14}
\end{align}
Hence, assuming they are independent, the probabilities in the first row are Eq.~\eqref{eq4.14} exponentiated to the number of PDCs needed in each protocol. The non-encoded protocol requires $\lceil (n+1)/2\rceil$ PDCs, while our protocol requires $2n+1$ PDCs.

The second row comes from the probability of getting a four-GHZ state from a PBS when inputting two Bell states into it. The transformation is as follows
\begin{align}\label{eq4.15}
\ket[1234]{\Phi_1,\Phi_1} \xrightarrow[]{\text{PBS}} &\frac{1}{2}[\ket{V;HV;0;H}-\ket{H;V;V;H}\\
&-\ket{V;H;H;V}+\ket{H;0;HV;V}]_{1 2^\prime 3^\prime 4}. \nonumber
\end{align}
We can see that there is a probability of $\frac12$ of getting one photon in each mode ($2'$ and $3'$), which is the GHZ state up to local operations. Assuming each GHZ state creation process is independent, the total probability is the product from every PBS needed. Hence, in the second row, it is $1/2$ exponentiated to the number of PBSs needed. Our protocol does not required PBS, while the existing non-encoded protocol uses $\lfloor n/2\rfloor$ ones.

The second row comes from the success probability of unitary gates. Most of them are passive optical devices, such as, \textsf{BS} and phase shifters. There are also \textsf{CNOT} gates, constructed from fusion gates. Our encoded protocol uses these gates to encode logical qubits and to construct its error correction facilities. The existing protocol does not need the above processes, hence, requires none of these gates. 

We will now consider the success probability of each gate one-by-one. Let $\epsilon_\textsf{G}<1$ be the probability when gate \textsf{G} is successful, where $\textsf{G} \in \{\textsf{U},\textsf{H},\textsf{H}_\text{log},\textsf{CNOT}\}$. Assuming each gate is independent of the others, the probability for all gates to be successful is the product of every $\epsilon_\textsf{G}$.

Generally, gate \textsf{U} consists of phase shifters and beam splitters, while gate \textsf{H} is made up of a 50-50 beam splitter. Since they are made up of passive devices, we will assume that these two one-photon gates will succeed in almost every events, i.e., $\epsilon_\textsf{U} \rightarrow 1$ and $\epsilon_\textsf{H} \rightarrow 1$. It is considered a reasonable assumption, verified by a real experiment, since $(1-\epsilon_\textsf{G}) \sim 10^{-9} - 10^{-12}$ \cite{proietti2020experimental}.

The action of gate $\textsf{H}_\text{log}$ can be written as
\begin{align}
\textsf{H}_\text{log} = \frac{1}{\sqrt{2}}(\textsf{X}_\text{log}+\textsf{Z}_\text{log}),
\label{eq4.16}
\end{align}
where, $\textsf{X}_\text{log}$ and $\textsf{Z}_\text{log}$ are logical \textsf{X} gate and logical \textsf{Z} gate, respectively. The $\textsf{X}_\text{log}$ gate is constructed from a sequence of one-qubit \textsf{X}s, and similarly, $\textsf{Z}_\text{log}$ gate is constructed from a sequence of one-qubit \textsf{Z}s. This suggests that the two logical gates are combinations of phase shifters and beam splitters, which are passive devices. Hence, we will assume that their success probabilities $\epsilon_{\textsf{X}_\text{log}} \rightarrow 1$ and $\epsilon_{\textsf{Z}_\text{log}} \rightarrow 1$, which leads to $\epsilon_{\textsf{H}_\text{log}} \rightarrow 1$, too.

The \textsf{CNOT} gate for parity encoded qubits was used in Sec.~\ref{sec:nehurijwp}. It consists of two fusion gates, $\mathcal{F}_I$ and $\mathcal{F}^\prime_{II}$ and two $\textsf{X}_\text{log}$ gates. We have already discussed $\textsf{X}_\text{log}$ to have $\epsilon_{\textsf{X}_\text{log}} \rightarrow 1$. Next we consider fusion gates $\mathcal{F}_I$ and $\mathcal{F}^\prime_{II}$. Each gate consists of one PBS, hence, the success probability is $\frac12$ per gate. The success probability of a \textsf{CNOT} gate is then given by $\epsilon_{\textsf{CNOT}}$
\begin{align}
\epsilon_{\textsf{CNOT}} = \frac{1}{2}\cdot\frac{1}{2} = \frac{1}{4}.
\label{eq4.17}
\end{align}Hence, $2n$ \textsf{CNOT} gates have success probability of
\begin{align}
p(\text{gates}) = \left(\frac{1}{4}\right)^{2n}.
\label{eq4.18}
\end{align}The only contribution of non-unity probability from quantum gates is from \textsf{CNOT} gates. Other gates, including \textsf{U},\textsf{H} and $\textsf{H}_\text{log}$ are assumed to have unit probability of success. Hence, the bottom right cell of Table 3 is given by Eq.~\eqref{eq4.18}.

\begin{figure}[t!]
\centering
\includegraphics[width=\columnwidth]{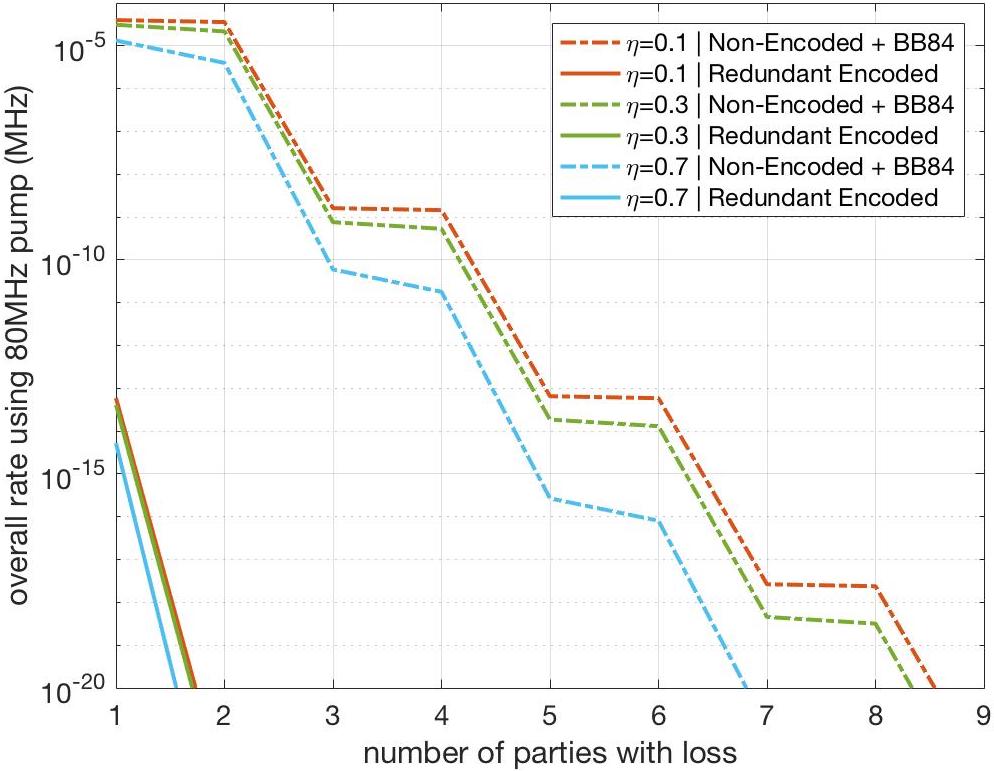}
\caption{Rate of creating successful entanglement in MHz when using 80 MHz pump rate for $\eta = 0.1, 0.3$ and 0.7 for both non-encoded with BB84 protocol and the redundant encoded protocol.}
\label{p-rateN}
\end{figure}

\begin{figure}[t!]
\centering
\includegraphics[width=\columnwidth]{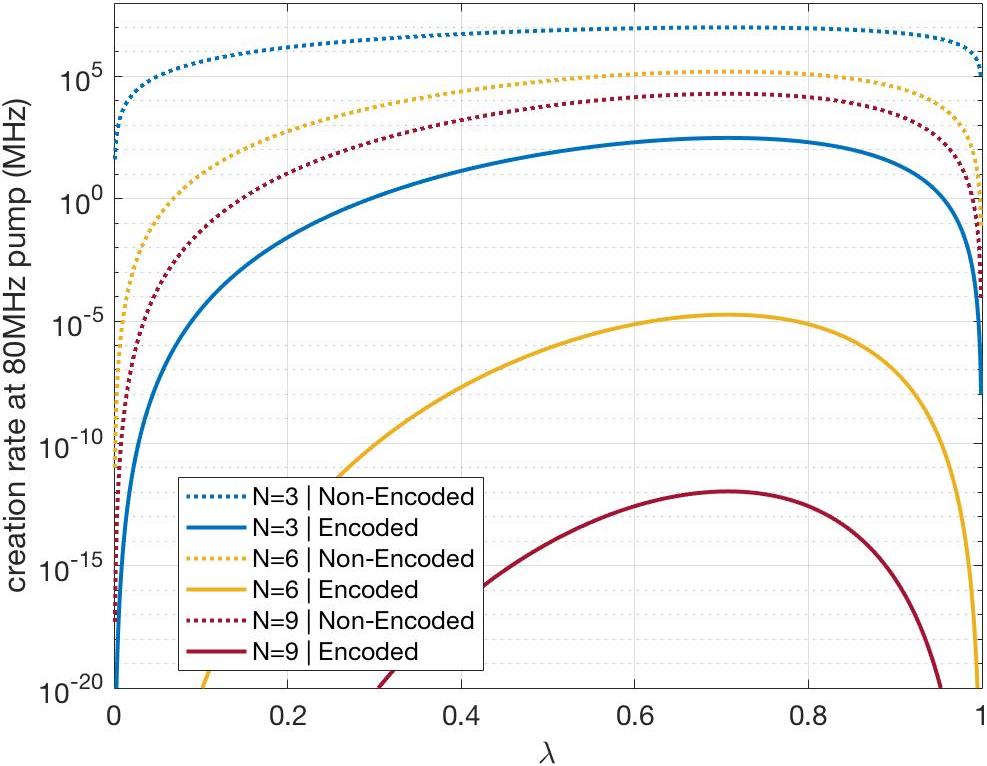}
\caption{Rate of successful entanglement creation as a function of $\lambda$ using 80 MHz pump, where $p = \lambda^2 (1-\lambda^2)$ is the probability of getting a Bell state from a PDC of different $\lambda$ for three ($N=3$), six ($N=6$) and nine ($N=9$) parties entanglement distribution using non-encoded and 2-photon redundant encoded protocols.}
\label{p-ld}
\end{figure}

Combining the result from Table~\ref{tab:poretjngkf}, the probability for creating one GHZ state for the CKA protocol implemented by Proietti \emph{et al.}~\cite{proietti2020experimental}---shown in Fig.~\ref{setup}---is
\begin{align}
 p_{C,\text{non-enc}} &= \left(\lambda^2(1-\lambda^2)\right)^{\lceil (n+1)/2\rceil} \cdot (\frac{1}{2})^{\lfloor n/2\rfloor}.
\label{eq4.19}
\end{align}
The probability for creating one encoded GHZ state for our presented protocol is
\begin{align}
 p_{C,\text{enc}} &= \left(\lambda^2(1-\lambda^2)\right)^{2n+1} \cdot \left(\frac{1}{4}\right)^{2n}.
\label{eq4.20}
\end{align}
Assuming the photon pump is running at 80 MHz, we can find rate of entangled state creation by 
\begin{align}
 \text{rate} = p_C \times \text{pump rate}.
 \label{eq4.21}
\end{align}
To illustrate, Fig.~\ref{p-rateN} shows the creation rate using the probabilities from Eq.~\eqref{eq4.19} and Eq.~\eqref{eq4.20} on a log scale for $\eta = 0.1, 0.3$ and 0.7. We can see that the rate for the encoded protocol decreases substantially compared to the non-encoded protocol. This is mainly due to difference in the exponent $N/2$ and $2N$ on $\lambda^2 (1-\lambda^2) \sim 10^{-4}$ in Table~\ref{tab:poretjngkf}. Unfortunately, $p = \lambda^2(1-\lambda^2) \ll 1$ for real $\lambda$. Hence, it is unlikely that the encoded protocol can perform better than the non-encoded protocol if the photon creating device is PDC without further resources such as multiplexing. In Fig.~\ref{p-ld} we show the dependence of the creation rate (MHz) on $\lambda$. However, note that increasing $\lambda$ comes at the cost of creating multiple pairs in the PDCs, which will cause spurious detection events and a severe degradation of the GHZ states. Our analysis is valid only when $\lambda\ll1$.

\begin{figure}[t!]
\centering
\includegraphics[width=\columnwidth]{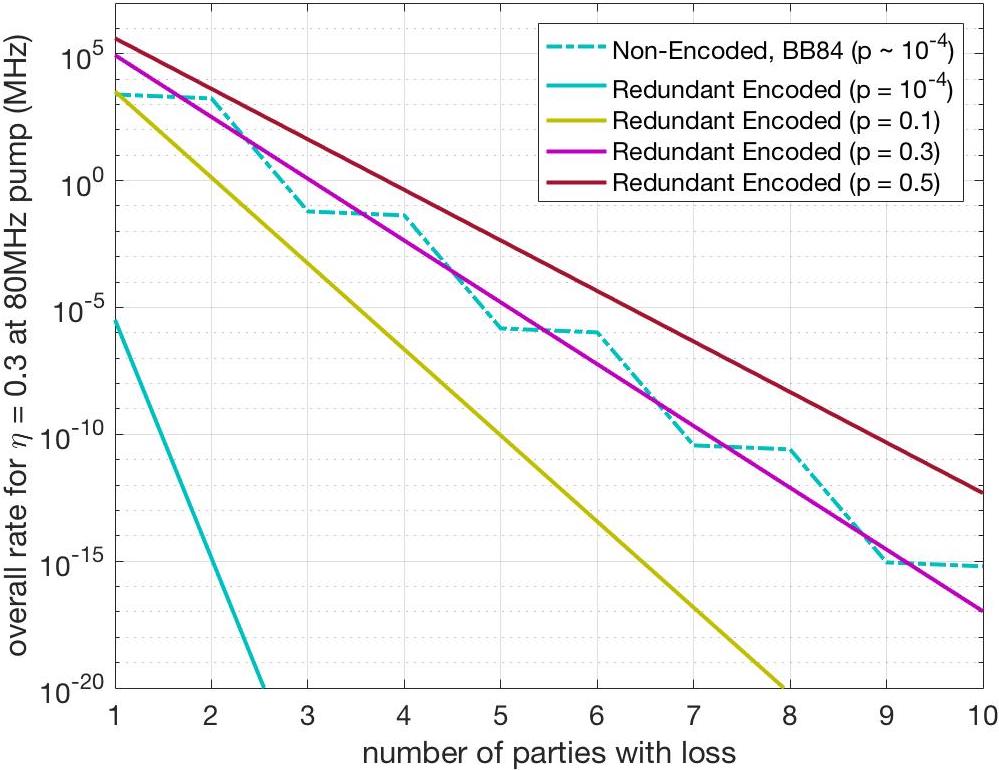}
\caption{Entanglement distribution rate when $\eta = 0.3$ using 80 MHz pump for PDC, $p \sim 10^{-4}$, and three other imaginary devices with $p = 0.1,0.3$ and 0.5.}
\label{p-rateN-p}
\end{figure}

Since the redundant encoded protocol will not be useful with PDC as photon creating device, we may consider using other devices with higher probability of creating photons than PDC. Fig.~\ref{p-rateN-p} plots the overall entanglement distribution rate as a function of number of parties for different values of $p$. It is promising that, with $p \sim 0.3$, the performance of both protocols are comparable. With higher $p$, the redundant encoded protocol can perform better than the non-encoded protocol. Hence, if we have access to sources with high probability of producing photons, the exponential drop in probability due to the photon creation rate in PDC could, in principle, be recovered.

It is reasonable to be hopeful for devices with photon creation probability higher than PDC to be more widely available in the foreseeable future. At the moment, there are other sources of single and entangled photons. For single photons, there are, e.g., trapped ions \cite{duan2004scalable}, cold atoms \cite{Lounis_2005} and colloidal CdSe/ZnS quantum dots \cite{Brokmann_2004}. For entangled photons, there are, e.g., atomic ensemble \cite{Kuzmich_2003} and biexciton-exciton cascade quantum dots \cite{QDsolidstate}. Although most of these devices are still in the experimental stage, they are evolving rapidly, hence, are good candidates for the desired photon sources.


\section{Discussion and Conclusions}\label{sec:gbehirudfj}

Conference key agreement (CKA) is an information processing task where more than two parties want to share a common secret key. The form of the protocol that uses GHZ states suffers from extreme sensitivity to photon loss.  Here, we introduced a redundantly encoded protocol with error correction for CKA, making the protocol resilient to photon loss both in the detector and the transmission line. We assume each part has the same loss parameter, but this is easily generalised to parties with different loss parameters. 

We compare the performance of our protocol and the existing protocol in terms of their rates of creating and transmitting entangled states. Our protocol provides a speed-up in transmission rate over the existing protocol. However, extra cost is required for encoding and error correction in our protocol. Using parametric downconverter (PDC) as photon sources, the extra cost for the protocol quickly becomes too high to be implemented. We showed that, using entangled photon sources with creation probability $p \gtrsim 0.3$, the loss-tolerant protocol can outperform the original CKA protocol. This is much higher than what PDC can provide. Its secret key rate also overcomes the existing protocol's rate. Hence, the loss-resilient CKA protocol presented here requires high probability entangled photon sources. Although these devices have not yet been widely distributed commercially, they are currently in an experimental stage. Promising candidates are entangled photon sources from atomic ensemble \cite{Kuzmich_2003} and biexciton-exciton cascade quantum dots \cite{QDsolidstate}.

Our error correction protocol requires quantum-nondemolition detectors (QND), which are experimentally very challenging. There are in principle many different ways to implement QND, e.g., using physical processes, i.e., cross-Kerr nonlinearities \cite{inbook,Sagona_Stophel_2020}, photon-cavity interactions \cite{Guerlin_2007,Nogues1999}, and detecting photons interferometrically using linear optics \cite{Kok_2002}. Single photon resolution is still challenging using cross-Kerr nonlinearities \cite{inbook,PhysRevLett.57.2473}. However, with cavity quantum electrodynamics and interferometry in linear optics, they have successfully realised the single photon resolution on QND \cite{birnbaum2005photon,Wilk488,Kok_2002}. 

It is an open question whether our loss-tolerant protocol can be implemented with lower complexity. Different ways of encoding qubits could be explored to reduce the entanglement creation cost. If a certain reduction of complexity is achieved, we might be able to implement loss-resilient CKA protocol without having to wait for experimental devices such as QND detectors.

\section*{Acknowledgements}\noindent
The authors acknowledge the support of EPSRC via the Quantum Communications Hub through grant number EP/M013472/1.

\end{document}